\documentclass[trackchanges]{aastex7}
\DeclareUnicodeCharacter{02BC}{\textquotesingle}

\usepackage{graphicx}
\usepackage{subcaption}
\usepackage{amsmath}
\usepackage{bm}

\begin{document}

\title{Beyond Colors: Probing Redshifts from Galaxy Morphology in Single-band Images with ViT-MDNz}

\author[orcid=0009-0009-1617-8747,sname=Luo]{Zhijian Luo}
\affiliation{Shanghai Key Lab for Astrophysics, Shanghai Normal University, Shanghai 200234, People’s Republic of China}
\email{zjluo@shnu.edu.cn} 

\author{Yangyang Li}
\affiliation{Shanghai Key Lab for Astrophysics, Shanghai Normal University, Shanghai 200234, People’s Republic of China}
\email{liyangyang11223@163.com}  

\author[orcid=0009-0005-1876-699X, gname=Jianzhen, sname=Chen]{Jianzhen Chen} 
\affiliation{Shanghai Key Lab for Astrophysics, Shanghai Normal University, Shanghai 200234, People’s Republic of China}
\email[show]{jzchen@shnu.edu.cn}

\author[gname=Qishen, sname=Cao]{Qishen Cao} 
\affiliation{Shanghai Key Lab for Astrophysics, Shanghai Normal University, Shanghai 200234, People’s Republic of China}
\email{1000588518@smail.shnu.edu.cn}

\author[sname=Cao]{Duo Cao}
\affiliation{Joint Lab for Submillimeter Astronomy, Shanghai Normal University, Shanghai 200234, People's Republic of China}
\email[show]{dcao@shnu.edu.cn}

\author[0000-0002-2326-0476,sname=Zhang,gname=Shaohua]{Shaohua Zhang}
\affiliation{Shanghai Key Lab for Astrophysics, Shanghai Normal University, Shanghai 200234, People’s Republic of China}
\email{zhangshaohua@shnu.edu.cn}

\author[0000-0001-8244-1229,sname=Xiao,gname=Hubing]{Hubing Xiao}
\affiliation{Shanghai Key Lab for Astrophysics, Shanghai Normal University, Shanghai 200234, People’s Republic of China}
\email{hubing.xiao@outlook.com}

\author[gname=Chenggang]{Chenggang Shu}
\affiliation{Shanghai Key Lab for Astrophysics, Shanghai Normal University, Shanghai 200234, People’s Republic of China}
\email{cgshu@shao.ac.cn}

\begin{abstract}

To address the challenge of estimating redshifts when only single-band images are available, this study introduces a deep learning model named ViT-MDNz. Leveraging robust statistical priors learned from large-scale data concerning the correlation between redshift and morphology, the model can directly estimate redshifts and their associated uncertainties from single-band galaxy images. It integrates a Vision Transformer (ViT) to extract deep morphological features and a Mixture Density Network (MDN) to predict the full redshift probability density function. Trained and evaluated on approximately 300,000 single-band images from the DESI Legacy Imaging Surveys (DESI-LS), the model achieves a normalized median absolute deviation $\sigma_{\rm NMAD} = 0.034$ and an outlier fraction $f_{\rm out} = 2.6\%$ in the $r$‑band for redshifts up to $z \lesssim 1$. Evaluations using probability integral transform (PIT) and continuous ranked probability score (CRPS) confirm that the predicted probability density functions are well calibrated and closely match the true distribution. These results demonstrate that competitive redshift estimates can be obtained using morphological features alone, and that incorporating color information further enhances the accuracy and robustness of the estimation. Therefore, ViT-MDNz provides a practical approach for redshift estimation of galaxy samples with limited photometric band coverage, contributing to improved completeness and usability of redshift catalogs for future large-scale surveys such as DESI and LSST.


\end{abstract}



\section{Introduction} \label{sec:intro}

Galaxy photometric redshift estimation is a crucial technique that leverages multi-band photometric data, combined with template fitting or machine learning methods, to infer the cosmological redshift of galaxies. This approach significantly alleviates the reliance on time-consuming spectroscopic observations for every individual target, enabling the efficient acquisition of three-dimensional spatial distribution information for large samples of galaxies. It has become a cornerstone of modern cosmological surveys for obtaining vast amounts of redshift data. Currently, nearly all imaging-dominated survey projects—including completed ones like the Sloan Digital Sky Survey (SDSS; \citealt{1996AJ....111.1748F,2000AJ....120.1579Y}), the Dark Energy Survey (DES; \citealt{2016MNRAS.460.1270D,abbott2021dark}), and the Kilo-Degree Survey (KiDS; \citealt{2013ExA....35...25D}), as well as ongoing and future missions such as the Euclid Space Telescope \citep{laureijs2011euclid}, the Wide Field Infrared Survey Telescope or Nancy Grace Roman Space Telescope (WFIRST; \citealt{2012arXiv1208.4012G,spergel2015widefield,2019arXiv190205569A}), and the
Legacy Survey of Space and Time (LSST; \citealt{abell2009lsst,2019ApJ...873..111I})—heavily rely on photometric redshifts to derive redshift information for millions to billions of celestial objects, thereby supporting key scientific investigations into large-scale structure and dark energy.

However, the accuracy and reliability of existing photometric redshift methods, whether based on spectral energy distribution (SED) template fitting or machine learning regression, are highly dependent on multi-band photometric data. Multiple bands collectively characterize key spectral features in a galaxy's SED, such as the 4000Å break and the Lyman break, providing the essential color information for redshift estimation. Mainstream template-fitting tools like HyperZ \citep{2000A&A...363..476B}, EAZY \citep{2008ApJ...686.1503B}, and LePhare \citep{1999MNRAS.310..540A} typically require input from no fewer than four photometric bands to ensure reliable results. Similarly, machine learning methods—both traditional algorithms like TPZ  \citep{2013MNRAS.432.1483C}, ANNz2 \citep{2016PASP..128j4502S}, and Random Forests \citep{2013MNRAS.432.1483C,2015MNRAS.452.3710R,2021MNRAS.502.2770M,2024MNRAS.52712140L}, and more recent deep learning models such as Delight \citep{2017ApJ...838....5L}, CNNs \citep{2015MNRAS.452.4183H,2018A&A...609A.111D,2019A&A...621A..26P,2021ApJ...909...53Z}, BNNs \citep{zhou2022photometricBNN}, and LSTMs \citep{2024MNRAS.535.1844L}—generally rely on data from more than four bands during both training and inference. When the number of available bands drops below four, the severe lack of color information leads to a significant degradation in redshift estimation accuracy.

Furthermore, in practical observations, factors such as observational constraints, data gaps, or contamination from foreground/background sources often prevent the acquisition of complete multi-band data. In extreme cases, only single-band imaging may be available, which severely undermines the accuracy and reliability of photometric redshift estimates—or even renders them infeasible. Current mitigation strategies generally fall into two categories: first, performing Bayesian imputation or generative inpainting for partially missing bands before applying conventional redshift estimation methods (see, e.g., \citealt{2024MNRAS.531.3539L}). Although this approach can handle incomplete data, it introduces additional systematic errors that grow with the number of missing bands. Second, discarding sources that lack key bands or have low signal-to-noise ratios altogether. While this avoids imputation-related errors, it sacrifices valuable sample size, and—because data missingness is often non-random—it may introduce selection bias, thereby compromising the reliability of subsequent cosmological statistical inferences.


Faced with these challenges, a fundamental question arises: is it still possible to obtain reasonably reliable redshift estimates when color information is extremely limited—for instance, when only single-band images are available? It should be noted that single-band images lack sufficient information to fully break the degeneracy between redshift, galaxy type, and luminosity, and thus cannot achieve “precise measurement” in the traditional sense. The goal of this study is not a breakthrough in physical principles, but rather to explore viable approaches for redshift estimation under strict observational constraints by leveraging statistical priors learned from large-scale data.

We argue that although single-band images lack direct color information, they may still contain morphological and structural features that are statistically correlated with redshift. In large-scale cosmological simulations and observations, there exists a statistical correlation between the morphological characteristics of galaxies—such as angular size, concentration index, surface brightness profiles—and their intrinsic physical properties as well as redshift. For instance, high-redshift galaxies are generally smaller and more irregular in shape, while certain typical morphologies—such as massive elliptical galaxies or galaxies with well-defined spiral arms—are more commonly observed at low redshifts. As redshift increases, the angular diameter of galaxies tends to decrease, their morphology becomes more compact, and cosmological effects systematically alter their apparent brightness and shape profiles.

However, redshift signals that rely on morphology are often very weak and difficult to extract effectively using handcrafted features. Therefore, this study introduces feature extractors capable of automatically learning complex spatial patterns, combined with training on large-scale high-quality samples. By learning from a large number of single-band images, the model essentially constructs a joint probability distribution between morphological features and redshift, thereby providing new observational targets with redshift probability estimates based on statistical priors. Although this method cannot fully replace multi-band photometry or spectroscopic observations, it offers a statistically reliable and scientifically valuable complementary path for redshift estimation under severely data-limited conditions.

Based on this, this study is the first to introduce the vision transformer (ViT) architecture into the field of photometric redshift estimation, constructing a hybrid model named ViT‑MDNz. The innovation of this model is mainly reflected in two aspects: (1) leveraging the global attention mechanism of ViT to directly extract deep morphological features related to redshift from single-band images, without relying on multi-band color information; (2) feeding the extracted features into a mixture density network (MDN) to predict the complete redshift probability density function (PDF), rather than just a single point estimate. This approach not only provides accurate point estimates, but also quantifies the uncertainty of the predictions.

We validated the method using large-sample data from the DESI Legacy Imaging Surveys (DESI-LS; \citealt{2021AAS...23723503S}). Experimental results show that even under the extreme condition of using only single-band images as input, ViT‑MDNz can still achieve competitive redshift estimation accuracy and reliable distribution prediction performance. This technology provides a novel and effective solution for handling galaxy samples with insufficient band coverage due to observational constraints in current and future survey projects (such as LSST, Euclid), and is expected to significantly improve the completeness of redshift catalogs and the efficiency of data utilization.

The structure of this paper is as follows: Section \ref{sec:dataset} describes the datasets used and the preprocessing pipeline; Section \ref{sec:method} details the architecture design and training strategy of the ViT-MDNz model; Section \ref{sec:performance} presents and analyzes the experimental results for both single-band and multi-band scenarios; Section \ref{sec:summary} summarizes the work and discusses future research directions.

\section{ Data and Preprocessing} \label{sec:dataset}

The primary objective of this study is to develop and validate a deep learning model capable of estimating redshifts directly from single-band galaxy images. To achieve this goal, we utilized large-sample data from the DESI-LS and implemented rigorous data screening and preprocessing procedures to ensure data quality and the effectiveness of model training.

\subsection{Data Source and Sample Construction}

The galaxy image data used in this study originate from the ninth data release (DR9) of the DESI-LS. As a key project that provides imaging data for the Dark Energy Spectroscopic Instrument (DESI) \citep{2019AJ....157..168D}, DESI-LS aims to supply galaxy and quasar targets for subsequent DESI observations to support its cosmological studies. DESI is installed on the 4-meter Mayall Telescope at Kitt Peak National Observatory, while DESI-LS comprises three complementary imaging surveys: the Beijing-Arizona Sky Survey (BASS), the Mayall $z$-band Legacy Survey (MzLS), and the Dark Energy Camera Legacy Survey (DECaLS).

BASS and MzLS collectively cover the Northern sky. BASS uses the 2.3-meter Bok Telescope at Kitt Peak to obtain $g$- and $r$-band data \citep{2004SPIE.5492..787W}, while MzLS utilizes the 4-meter Mayall Telescope (hosting DESI) for $z$-band observations \citep{2016SPIE.9908E..2CD}; these are jointly referred to as BASS/MzLS. DECaLS, conducted using the 4-meter Blanco Telescope at the Cerro Tololo Inter-American Observatory in Chile, images the Southern sky in the $g$, $r$, and $z$ bands \citep{2015AJ....150..150F}. DECaLS, combined with BASS/MzLS, provides three-band ($g$, $r$, $z$) observational data covering approximately 14,000 square degrees. Additionally, the Dark Energy Survey (DES) \citep{2016MNRAS.460.1270D}), also using the DECam instrument on the Blanco Telescope, contributed $g$, $r$, $z$ band data covering 5,000 square degrees in the South Galactic Cap, which is also included in the DESI-LS data release.

To obtain reliable spectroscopic redshifts as ground truth for training and testing, we cross-matched the DESI-LS DR9 photometric catalog with the spectroscopic catalog from the Sloan Digital Sky Survey (SDSS) Data Release 16 (DR16). The cross-matched catalog of DESI-LS DR9 and SDSS DR16 has been publicly released online\footnote{\url{https://portal.nersc.gov/cfs/cosmo/data/legacysurvey/dr9/north/external/}} \footnote{\url{https://portal.nersc.gov/cfs/cosmo/data/legacysurvey/dr9/south/external/}}, containing approximately 5.8 million objects. By matching the SDSS DR16 spectroscopic identifiers (MJD, PLATE, FIBERID) provided in this catalog, we obtained their spectroscopic redshifts from SDSS. Building upon this, we further filtered the sample to include only those with reliable spectroscopic quality ($z\_warning = 0$), and selected galaxies classified as REX (round, variable-axis-radius galaxies), DEV (de Vaucouleurs profile galaxies), EXP (exponential disk galaxies), and SER (Sérsic profile galaxies). Furthermore, we removed all objects with $maskbits !=$ 0, as these are either located in corrupted imaging pixels or near bright stars, globular clusters, or nearby galaxies. This process resulted in a high-quality sample of approximately 2.6 million galaxies, possessing both DESI-LS DR9 photometric information and precise SDSS DR16 spectroscopic redshifts.

\begin{figure} 
        \includegraphics[width=0.96\textwidth]{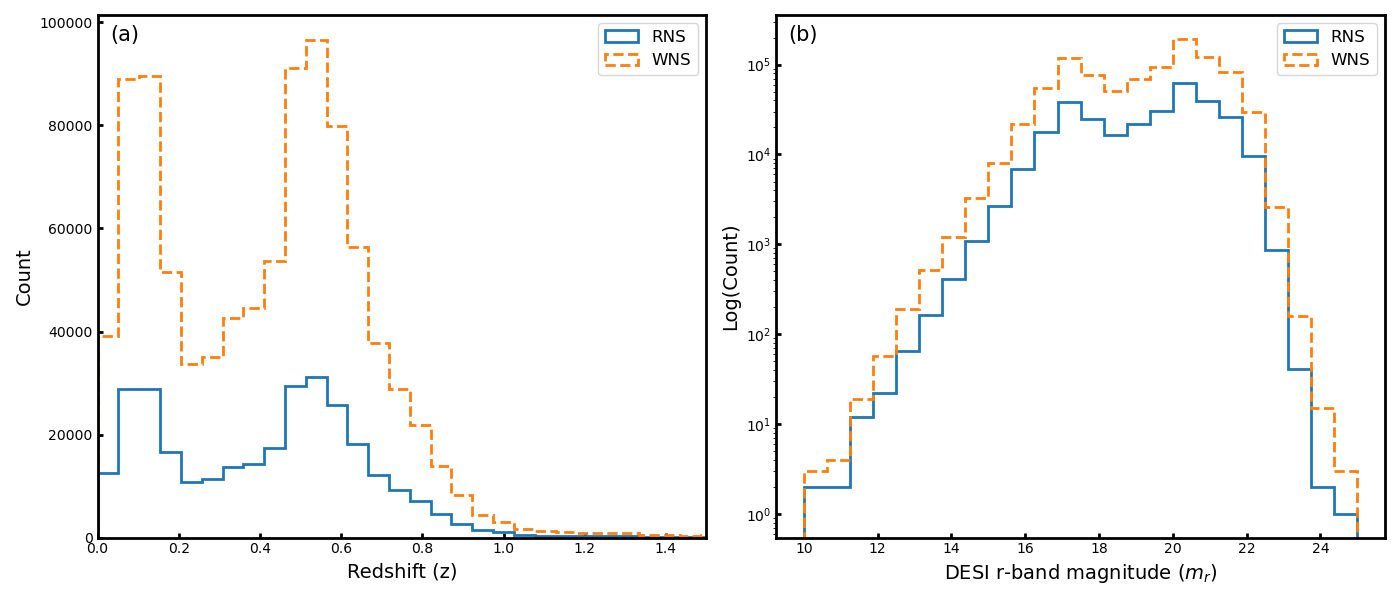} 
        \centering
        \caption{Comparison of the RNS and WNS samples: (a) number of galaxies as a function of spectroscopic redshift in linear scale, and (b) distribution of DESI $r$-band magnitudes.} 
        \label{fig:redshift} 
\end{figure}

Given the storage and computational challenges associated with directly processing the image data for all 2.6 million galaxies, we randomly selected a subset of 300,000 Northern sky galaxies from this sample for the present study to streamline the research process. We downloaded the corresponding $g$, $r$, and $z$ band photometric image cutouts for these galaxies from the image data released on the DESI Legacy Imaging Surveys website \footnote{\url{ https://www.legacysurvey.org/viewer/}}. Figure \ref{fig:redshift}(a) shows a comparison of the spectroscopic redshift distributions in linear scale between the 300,000-galaxy Random Northern Sample (RNS) used in this study and the complete Whole Northern Sample (WNS). Additionally, Figure \ref{fig:redshift}(b) compares the distributions of their DESI $r$-band magnitudes. The distributions in both panels are highly consistent, confirming that the random subset remains statistically representative of the parent spectroscopic catalog. It is important to emphasize, however, that this parent catalog is highly  non-representative of the underlying total galaxy population. It consists of galaxies from the SDSS spectroscopic programs, including the main galaxy sample (MGS) and luminous red galaxies (LRGs), which are defined by specific selection functions and flux limits. Consequently, the redshift distribution exhibits a distinct bimodal structure and is significantly skewed, with approximately 99\% of the sample concentrated at $z \lesssim 1$. While our results demonstrate the model's success in learning the morphology-redshift relation for these specific populations, we acknowledge that applying the model to a truly representative or much deeper cosmic sample would require additional training with more complete datasets. 

\subsection{Data Preprocessing} \label{subsec:data_pre}

To ensure effective model learning and enhance training stability, we implemented a standardized preprocessing pipeline for all image data, primarily consisting of image cropping and pixel normalization.

First, during the image preprocessing stage, we performed center cropping on all galaxy images. The original images downloaded from DESI-LS have dimensions of 256×256 pixels with a spatial resolution of 0.262 arcseconds per pixel, with the target galaxy centered in each image. To preserve key morphological features while improving computational efficiency, all images were uniformly center-cropped to 192×192 pixels. This operation reduces the input data volume by approximately 44\%, significantly decreasing computational and memory overhead during model training and inference. Previous studies \citep{radford2015unsupervised,2025ApJS..277...22L,2025ApJS..279...17L,2025ApJS..280...11C,2025ApJS..280...69L} have shown that such cropping preserves the core structural information for centered astronomical targets and has been widely adopted in astronomical object recognition and classification tasks.

Second, regarding pixel normalization, the original pixel values range from [0, 255]. To promote training stability and convergence speed, we linearly normalized the pixel values to the range [-1, 1] using the following formula:
\begin{equation}
     x^* = \frac{x - 127.5}{127.5},
     \label{eq:normal} 
\end{equation}
where $x$ represents the original pixel value and 127.5 is the median of the original range. This normalization approach helps stabilize gradient flow during training and represents a standard practice in deep learning computer vision tasks \citep{patro2015normalization,radford2015unsupervised,goodfellow2016deep,he2016deep}.

Finally, for dataset splitting, we randomly divided the final sample of 300,000 galaxies into a training set (80\%) and a test set (20\%). The training set was used for learning model parameters, while the test set was reserved for the final evaluation of the model's generalization performance.

\section{Methodology} \label{sec:method}

To predict redshifts directly from single-band galaxy images, we propose a hybrid architecture named ViT-MDNz. This framework integrates a vision transformer (ViT) with a mixture density network (MDN), responsible for image feature extraction and photometric redshift probability density distribution modeling, respectively. The overall architecture of the ViT-MDNz model is illustrated in Figure \ref{fig:framework}. Detailed descriptions of the ViT and MDN modules are provided below. 

\subsection{ViT Encoder}

The ViT is a groundbreaking image processing model introduced by \citet{2020arXiv201011929D}. Its core innovation lies in completely abandoning traditional convolution operations and directly applying the standard transformer architecture, widely used in natural language processing, to sequences of image patches. This paradigm shift enables the model to capture global dependencies starting from the first layer, unlike convolutional neural networks (CNNs), which begin with local features and gradually expand their receptive field through multiple stacked layers.

\begin{figure} 
        \includegraphics[width=0.96\textwidth]{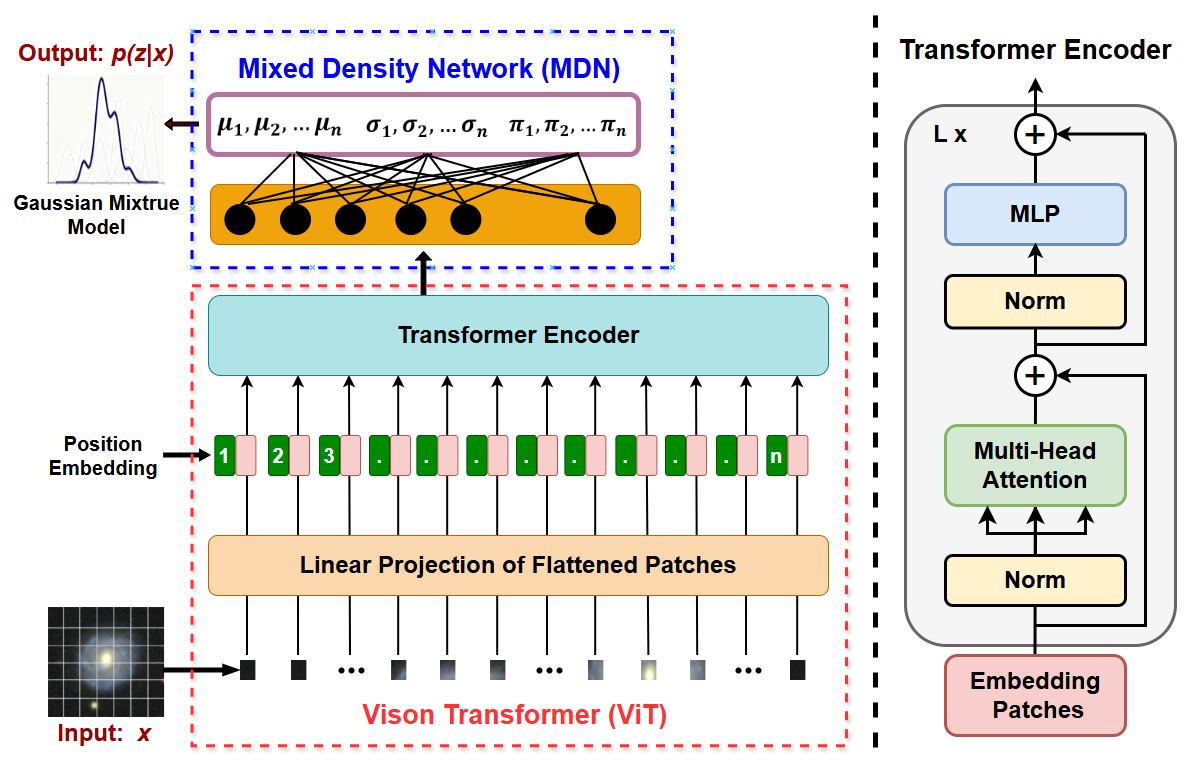} 
        \centering
        \caption{ViT-MDNz model architecture and data flow diagram. The input consists of single or multi-band galaxy images. These images are first processed by the ViT encoder (red dashed box): the image is split into fixed-size patches, which undergo linear projection and position encoding before being fed into the Transformer encoder layers to extract deep morphological features with global context. The extracted features are then passed to the MDN module (blue dashed box), which computes and outputs the conditional probability density function of the photometric redshift using a Gaussian Mixture Model.} 
        \label{fig:framework} 
\end{figure}

In our proposed ViT-MDNz model, the ViT encoder is responsible for extracting deep morphological features with global contextual information from the input galaxy images. Its primary processing pipeline is shown in the red dashed box in Figure \ref{fig:framework} and involves the following specific steps:

First, the input single-band galaxy image, denoted as $\mathbf{x} \in \mathbb{R}^{H \times W \times C}$  with height $H=192$, width $W=192$, and number of channels $C=1$, is divided into $N$ non-overlapping patches of size $P\times P$. Setting $P=8$ yields $N = (H \times W) / P^2 = 576$ patches. Each patch $x_i \in \mathbb{R}^{P \times P \times C}$ is flattened into a 1D vector of length $P\times P\times C = 64$ and then mapped via a trainable linear projection layer to a $D=128$ dimensional latent space, producing the patch embedding sequence $E \in \mathbb{R}^{N \times D}$.


Next, to preserve the spatial structure of patches within the original 2D image, a learnable 1D positional encoding $E \in \mathbb{R}^{N \times D}$ is added to the embedding sequence:
\begin{equation}
V_0 = E + E_{\text{pos}}
\label{eq:pos_patch}
\end{equation}
where $V_0 \in \mathbb{R}^{N \times D}$ represents the resulting sequence with incorporated positional information.

Unlike the standard Vision Transformer design, we omit the prepended classification token. Instead, after encoding, global average pooling is applied over the sequence dimension (i.e., across patches) to aggregate information from all image regions into a compact feature representation. This approach not only simplifies the architecture but also promotes a more balanced integration of semantic features across the entire image, which is better suited for regression-based redshift estimation.


The sequence $V_0$ is then processed by a stack of 8 Transformer encoder layers. Each layer consists of a multi-head self-attention (MHSA) module with 4 heads and a feedforward network (FFN) with a hidden dimension of 256. Residual connections and layer normalization are used in each sub-layer to facilitate stable training and mitigate vanishing gradients.


After the final encoder layer, the output sequence $V_8 \in \mathbb{R}^{N \times D}$ is obtained. Applying global average pooling along the patch dimension yields the compact image representation $V \in \mathbb{R}^{D}$, which integrates global semantic information and serves as a discriminative feature vector for subsequent redshift probability distribution modeling.


Compared to CNNs, the ViT encoder offers distinct advantages for galaxy image analysis. While CNNs rely on local receptive fields and progressive hierarchical abstraction to capture long-range dependencies indirectly, ViT’s self-attention mechanism enables direct global interactions among all patches from the first layer. This allows ViT to model relationships across different structural components—such as between the galactic bulge and spiral arms or tidal tails—more effectively, leading to a richer characterization of galaxy morphology and evolutionary state. Additionally, the Transformer architecture is highly scalable, benefiting from increased model capacity and data volume. This scalability enables the ViT encoder to learn generalizable features from growing astronomical datasets, supporting not only more robust redshift estimation but also providing a strong foundation for future large-scale survey analysis.

\subsection{Mixture Density Network} \label{subsec:mdn}

After processing by the ViT encoder, each input galaxy image is transformed into a 128-dimensional global feature vector $V \in \mathbb{R}^{128}$. This vector effectively encapsulates the overall morphology and structural information of the galaxy. As shown in the blue dashed box in Figure \ref{fig:framework}, this feature vector is fed into the MDN module to estimate the probability distribution of the redshift.

A mixture density network (MDN) is a neural network architecture specifically designed for complex regression problems \citep{bishop1994mixture}. The core idea of the MDN is to learn and output the complete conditional probability distribution $p(z|x)$. This method naturally captures the inherent uncertainty in predictions and potential multi-modal characteristics.

MDN combines the powerful fitting capability of neural networks with the flexibility of probabilistic models. Specifically, its final layer does not output a prediction directly, but instead outputs a set of parameters defining a mixture probability distribution, such as a Gaussian mixture model (GMM). For a univariate GMM with $K$ components (this paper sets $K=15$), the MDN needs to predict three sets of parameters for each input: the mixture weights $\pi_k$, the means $\mu_k$, and the variances $\sigma_k^2$ of each Gaussian component. The mixture weights must be non-negative and sum to 1, the means determine the central location of each component, and the variances control the spread of the distributions. Therefore, the MDN output layer must provide $3K$ parameters. The final conditional probability distribution can be expressed as the following mixture form:

\begin{equation}
p(z \mid x) = \sum_{k=1}^{K} \pi_k(x) \mathcal{N}\left( z \mid \mu_k(x), \sigma_k^2(x) \right),
\label{eq:mdn_p}
\end{equation}
where $\mathcal{N}$ denotes the Gaussian probability density function. All parameters – the weights $\pi_k$, means $\mu_k$, and variances $\sigma_k^2$ – are functions of the input features $x$, learned by the neural network.

To ensure the parameters meet their constraints, the MDN applies specific activation functions to its output layer: the mixture weights $\pi_k$ are processed by a softmax function to ensure normalization, and the variances $\sigma_k^2$ use an exponential function to ensure they are positive. The training of the MDN is based on maximum likelihood estimation, optimized by minimizing the negative log-likelihood loss:
\begin{equation}
\mathcal{L} = -\log p(z \mid x).
\label{eq:mdn_loss}
\end{equation}

Due to its ability to output full probability distributions, MDN has garnered significant attention in astronomical photometric redshift estimation. For instance, \citet{2018A&A...609A.111D} first combined MDN with a CNN to estimate redshift PDFs directly from multi-band images. Subsequent research, such as \citet{2024AJ....168..244Z}, further improved MDN's performance in redshift probability distribution modeling by integrating multi-source data through cross-modal fusion techniques. \citet{2024A&C....4900886T} employed Legendre memory units (LMU) for sequence modeling of photometric data combined with MDN, achieving high-precision photometric redshifts and PDF estimates.

In this study, our proposed ViT-MDNz model adopts a cascaded architecture. The MDN module takes the deep semantic features $V(x)$ extracted by the ViT as input, rather than the raw pixel data, to learn the posterior probability density function $p(z|x)$ of the redshift. This design fully leverages the advantages of ViT in complex visual pattern recognition and the flexibility of MDN in probabilistic modeling, establishing a solid technical foundation for achieving high-precision redshift prediction with reliable uncertainty estimation.

\subsection{ViT-MDNz Model Training Process}

The ViT-MDNz model is trained using an end-to-end supervised learning approach, jointly optimizing all parameters in both the ViT encoder and the MDN by maximizing the likelihood of the observed data. The core training objective is to make the predicted redshift probability distribution as close as possible to the true distribution. We use the negative log-likelihood loss as the optimization target, defined as:
\begin{equation}
\mathcal{L} = -\log p(z_{\mathrm{spec}} \mid x) = -\log\left[ \sum_{k=1}^{K} \pi_k(x) \mathcal{N}(z_{\text{spec}} \mid \mu_k(x), \sigma_k^2(x)) \right]
\end{equation}
where $z_{spec}$ represents the true spectroscopic redshift of the galaxy, and $K=15$ is the number of Gaussian mixture components. Specifically, for each training sample, the model first extracts a 128-dimensional global feature vector via the ViT encoder. The MDN decoder then generates the mixture weights, means, and variances for the 15 Gaussian components. Finally, the negative log-likelihood of the probability distribution defined by these parameters at the true redshift value is calculated. By minimizing the sum of this loss over all training samples, the model gradually adjusts its parameters, causing the predicted distribution to become more concentrated near the true redshift.

Regarding the optimization strategy, we use the adam optimizer \citep{kingma2014adam} for parameter updates, with an initial learning rate set to $1\times10^{-4}$. The learning rate is decayed multiplicatively by a factor of $1/1.000004$ after each iteration. The training batch size is fixed at 64. The training process involves a complete forward and backward pass: input images are processed by the ViT encoder to extract features, the MDN outputs the probability distribution parameters, the loss is computed, and then all model parameters are updated uniformly via backpropagation.

The model was implemented using the TensorFlow framework with the Keras API. All experiments were performed on a computing platform equipped with an NVIDIA L40S GPU. Training consisted of 100,000 iterations per run, with no early stopping applied. This fixed iteration count was determined empirically to ensure robust convergence and training stability across all runs. On average, each experiment required approximately 3.6 GPU hours.



\section{Experiments and Results} \label{sec:performance}

This section presents a systematic evaluation of the proposed ViT-MDNz model's performance on the redshift estimation task. Experiments were conducted using the RNS dataset described in Section \ref{subsec:data_pre}. The dataset, comprising 300,000 galaxies, was partitioned into a training set (240,000 galaxies) and an independent test set (60,000 galaxies) in a 4:1 ratio. All models were trained exclusively on the training set, and their final performance metrics are reported on the held-out test set, ensuring an unbiased evaluation of generalization capability.

\subsection{Evaluation Metrics}

To comprehensively evaluate model performance, we employed two categories of standard metrics widely used in photometric redshift studies: point estimate metrics and probability distribution metrics. The former assesses the accuracy of the redshift point prediction value $z_{pred}$, while the latter evaluates the reliability and statistical calibration quality of the predicted PDF.

For the evaluation of redshift point estimates, we take the expectation of the conditional probability density distribution predicted by the model $p(z \mid x)$  as the point estimate value $z_{\rm{pred}}$, which is then compared with the true spectroscopic redshift $z_{\rm{spec}}$. For a mixture density model composed of $K$
Gaussian components, this estimate is given by:
\begin{equation}
z_{\text{pred}} = \mathbb{E}_{z \sim p(z \mid \mathbf{x})}[z] = \sum_{k=1}^{K} \pi_k(x) \, \mu_k(x),
\label{eq:z_pred_def}
\end{equation}
where $\mu_k(x)$ are the mixture weights for each component and $\mu_k(x)$ are the corresponding Gaussian means. The core evaluation metrics used include the normalized median absolute deviation ($\sigma_{\mathrm{NMAD}}$), bias ($bias$), catastrophic outlier fraction ($f_{\mathrm{out}}$), mean squared error (MSE), and mean absolute error (MAE).

Among these metrics, $\sigma_{\mathrm{NMAD}}$ is used to evaluate the scatter of the redshift estimates. It is robust to outliers and less sensitive than the standard deviation. It is defined as follows \citep{2008ApJ...686.1503B}:
\begin{equation}
\sigma_{\mathrm{NMAD}} = 1.48 \times \mathrm{median} \left( \frac{|\Delta z - \rm{median(\Delta z)}|}{1 + z_{\mathrm{spec}}} \right),
\label{eq:nmad_def}
\end{equation}
where $\Delta z = z_{\mathrm{pred}}-z_{\mathrm{spec}}$, and the coefficient 1.48 is a scaling factor such that the $\sigma_{\mathrm{NMAD}}$ is comparable to the standard deviation under a Gaussian distribution assumption.

Bias quantifies the systematic offset in redshift estimation, reflecting the overall tendency of predictions to be higher or lower than the true values. It is defined as:
\begin{equation}
bias = \mathrm{median} \left( \frac{z_{\mathrm{pred}} - z_{\mathrm{spec}}}{1 + z_{\mathrm{spec}}} \right).
\label{eq:bias_def}
\end{equation}
This metric helps identify whether there is an overall overestimation or underestimation.

The catastrophic outlier fraction ($f_{\mathrm{out}}$) measures the proportion of severely erroneous redshift estimates. Following common standards \citep{2018A&A...619A..14F,2020A&A...644A..31E}, a prediction is classified as an outlier if it satisfies:
\begin{equation}
\frac{|z_{\mathrm{pred}} - z_{\mathrm{spec}}|}{1 + z_{\mathrm{spec}}} > 0.15.
\label{eq:outlier_cond}
\end{equation}
The outlier fraction$f_{\mathrm{out}}$ is the proportion of such outliers in the total sample.

As classic metrics in regression analysis, MSE and MAE measure the average discrepancy between predicted and true values from the perspectives of squared error and absolute error, respectively. They are defined as:
\begin{equation}
\mathrm{MSE} = \frac{1}{N} \sum_{i=1}^{N} (z_{\mathrm{pred}}^i - z_{\mathrm{spec}}^i)^2,
\label{eq:z_mse}
\end{equation}
\begin{equation}
\mathrm{MAE} = \frac{1}{N} \sum_{i=1}^{N} |z_{\mathrm{pred}}^i - z_{\mathrm{pred}}^i|,
\label{eq:z_mae}
\end{equation}
where $N$ is the total sample size, and $z_{\mathrm{spec}}^i$ and $z_{\mathrm{pred}}^i$ represent the spectroscopic redshift true value and the photometric redshift estimate for the $i$-th galaxy, respectively.

For PDF evaluation, we use continuous ranked probability (CRPS) and probability  integral transform (PIT) to systematically assess the consistency between the predicted probability distribution and the true distribution \citep{2018A&A...609A.111D}. Furthermore, we introduce the quantile-quantile (Q-Q) plot as a visual diagnostic tool to evaluate how closely the cumulative PIT distribution follows a uniform distribution. CRPS is a proper scoring rule that measures the overall accuracy of probabilistic forecasts. It provides a comprehensive scalar score for predictive performance by directly comparing the predicted cumulative distribution function (CDF) against the step function of the observation. For a predictive CDF $F$ and a true observation $x$, CRPS is defined as:
\begin{equation}
\mathrm{CRPS}(F, x) = \int_{-\infty}^{+\infty} \left[ F(y) - \mathbf{1}_{\{y \geq x\}} \right]^2 dy,
\label{eq:crps_def}
\end{equation}
where $y$ is the variable of integration, and $\mathbf{1}_{\{y \geq x\}}$ is the indicator function (which equals 1 if $y \ge x$, and 0 otherwise). A lower CRPS value indicates that the predicted probability distribution is closer to the true distribution.

PIT is a fundamental tool for assessing the calibration quality of probabilistic forecasts \citep{dawid1984present}. For each celestial object, its PIT value is the value of the predictive CDF evaluated at the true redshift $z_{spec}$:
\begin{equation}
\mathrm{PIT} = \int_{-\infty}^{z_{\mathrm{spec}}} \mathrm{PDF}(z) dz,
\label{eq:pit_def}
\end{equation}
where $\mathrm{PDF}(z)$ is the predicted redshift probability density function. Theoretically, for a perfectly calibrated set of PDFs, the PIT values for all samples should follow a standard uniform distribution $U(0,1)$ \citep{dawid1984present}. The intuition behind this property is: if the PDF accurately describes the true uncertainty, then the true redshift value can be seen as a random draw from that distribution, and thus its cumulative probability should be uniformly distributed between 0 and 1.

The shape of the PIT histogram provides intuitive evidence for diagnosing systematic biases in the PDFs: a U-shaped histogram indicates that the predicted PDFs are under-dispersed, meaning their uncertainties are underestimated and the predictions are overconfident. An inverse U-shaped (i.e., peaked) histogram indicates that the predicted PDFs are over-dispersed, meaning their uncertainties are overestimated and the predictions are overly conservative. Furthermore, monotonically changing slopes in the PIT distribution often suggest the presence of systematic offset errors in the predictions \citep{2016arXiv160808016P}. Due to its strong diagnostic power, PIT has been widely used in redshift PDF validation studies \citep{2010MNRAS.406..881B,2018A&A...609A.111D,2020A&A...644A..31E,2021MNRAS.502.2770M}.

\subsection{Single-band Redshift Estimation Results}

We first evaluate the performance of the ViT-MDNz model using single-band images. Table \ref{tab:sband_metrics} summarizes the point‑estimation results of the model on the test sets of three independent bands:  $g$, $r$, and $z$. Given that the vast majority of galaxies in our training sample ($\sim~99\%$) have redshifts below $z\lesssim1$, our evaluation focuses primarily on the model’s predictive accuracy and robustness within this redshift range, thereby assessing its practical applicability for estimating redshifts of common low‑redshift galaxies.

\begin{table}[ht]
\centering
\caption{ViT-MDNz Single-band Photometric Redshift Estimation Performance}
\label{tab:sband_metrics}
\setlength{\tabcolsep}{8pt}
\renewcommand{\arraystretch}{1.2}
\begin{tabular}{cccccc}
\hline
        \textbf{Input Band} & $\bm{\sigma_{\rm NMAD}}$ & $\bm{f_{\rm out} (\%)}$ & $\bm{bias}$ & \textbf{MSE} & \textbf{MAE} \\
        \hline 
        DESI $g$-band & 0.052 & 5.7 & -0.0039 &  0.0140 & 0.075 \\
       DESI $r$-band &\textbf{0.034} &\textbf{2.6} &\textbf{-0.0012} &\textbf{0.0075} &\textbf{0.051} \\
       DESI $z$-band & 0.034 & 4.1 & -0.0052 & 0.0111 & 0.059 \\
\hline
\end{tabular}
\end{table}

Analysis of Table \ref{tab:sband_metrics} shows that using only single-band images, the ViT-MDNz model achieves competitive results across all bands. The $r$-band performance is the best, with a $\sigma_{\rm NMAD}$ of 0.034, a low catastrophic outlier rate of 2.6\%, and a bias close to zero. This is likely because the $r$-band most effectively balances the signal-to-noise ratio and the detectability of morphological features for the low-redshift galaxies that dominate our sample. Although the performance in the $g$ and $z$ bands is slightly inferior, it remains significantly better than random guessing, demonstrating that the model successfully extracts redshift-related information from single-band images.


To place our single-band results in context, we compare them with the benchmark color-based results from the official SDSS catalog produced by \cite{2016MNRAS.460.1371B}. To ensure a fair comparison, we cross-matched our sample with the \cite{2016MNRAS.460.1371B} catalog using the MJD, PLATE, and FIBERID identifiers, identifying 284,417 galaxies (approximately 95\% of our study sample) common to both. For this identical subset, the 5-band ($ugriz$) photometric method of \cite{2016MNRAS.460.1371B} achieves a precision of $\sigma_{\rm NMAD} = 0.022$, with an outlier fraction $f_{\rm out} = 0.029$ and a bias of $0.0004$. In comparison, our $r$-band ViT-MDNz achieves $\sigma_{\rm NMAD} = 0.034$. While our precision is slightly lower, it is noteworthy that our model utilizes only a single photometric band and morphological information, yet remains competitive with a traditional 5-band color-based approach across a broad redshift range of $z \lesssim 1.0$.

While 5-band color-based models inherently offer higher precision due to their better sampling of the spectral energy distribution, our results demonstrate that single-band images can still achieve an accuracy level that is statistically valuable for large-scale structure studies. This is particularly significant in scenarios where multi-band data are unavailable, incomplete, or corrupted. Our model effectively ``rescues'' such data by exploiting the morphology-redshift relation—a physical link that is often neglected by traditional empirical methods or SED-fitting approaches that rely solely on multi-band color information.

\begin{figure} 
    \centering
    \includegraphics[width=0.8\textwidth]{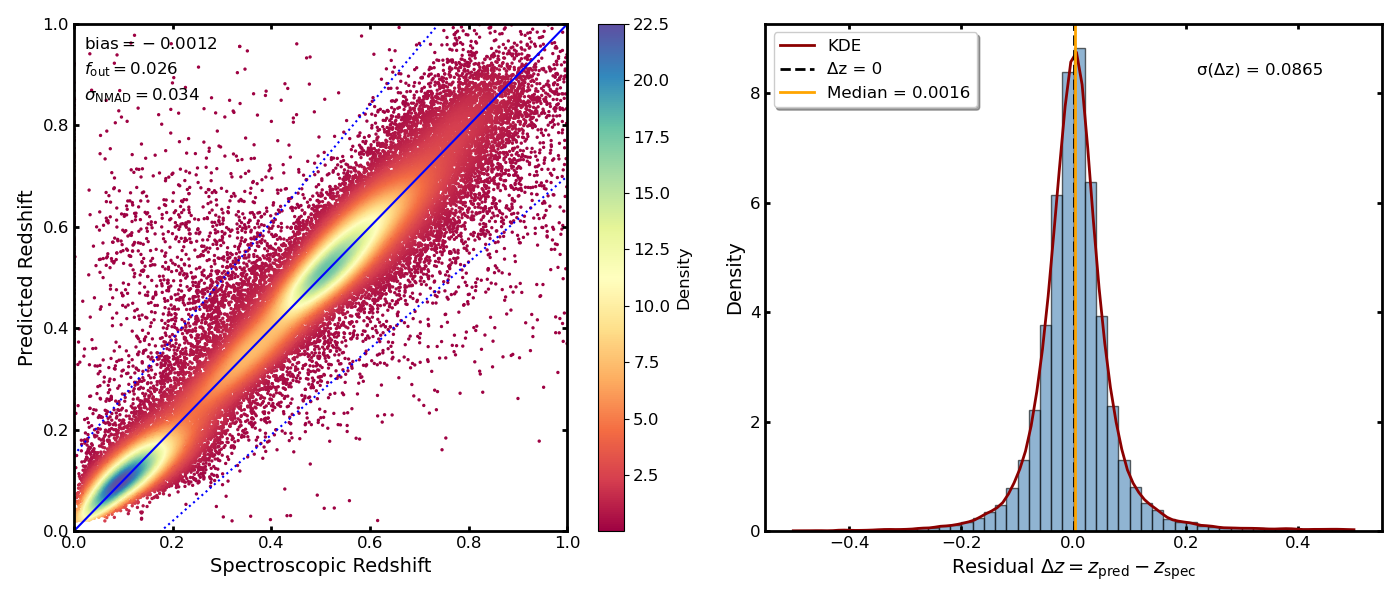} 
    \caption{(Left) Predicted redshift vs. spectroscopic redshift scatter plot for the r-band. (Right) Corresponding residual distribution. } 
    \label{fig:sing_zsp_zph} 
\end{figure}

Figure \ref{fig:sing_zsp_zph} visually illustrates the performance of the model in estimating photometric redshifts from the $r$-band. The density scatter plot on the left shows good agreement between the predicted redshifts and the spectroscopic redshifts, with data points densely clustered along the diagonal. This alignment is especially pronounced in the high-density regions around $z\sim 0.1$ and $z\sim 0.6$, indicating higher prediction accuracy at these typical redshifts. Although the data density decreases at other redshift intervals, the overall distribution of points still follows the diagonal trend, demonstrating robust performance across the entire redshift range of $z\lesssim1$. The residual plot on the right further supports this conclusion: the residuals are approximately symmetrically distributed around zero, with a mean (0.005) and median (0.0016) both close to zero, reflecting an overall unbiased prediction. The relatively small standard deviation of 0.0865 indicates that the prediction errors are concentrated, confirming high precision.

\begin{figure} 
    \centering
    \includegraphics[width=\textwidth]{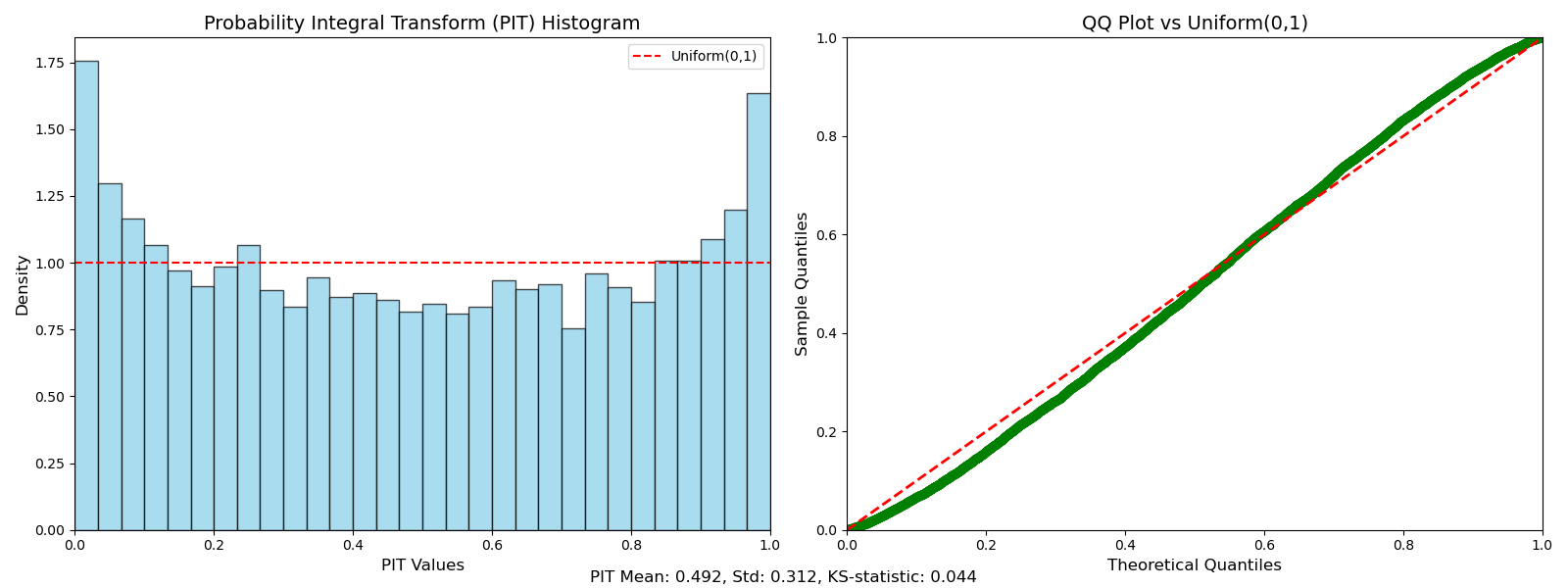} 
    \caption{(Left) PIT histogram for the r-band test set. The red dashed line indicates the ideal uniform distribution. (Right) Q-Q plot of the PIT values. Green points represent observed quantiles versus theoretical quantiles; the red dashed line is the y=x reference line.}
    \label{fig:pit} 
\end{figure}

A key strength of ViT‑MDNz lies in its ability to output a complete photometric redshift probability density function (PDF) for each galaxy. We generated the corresponding predicted redshift PDFs using single‑band images from the test set and assessed their reliability. To quantify the calibration quality of the PDFs, we computed the PIT values for the $r$‑band test set; the resulting distribution and Q‑Q comparison are shown in Fig.~\ref{fig:pit}. The PIT histogram is generally close to uniform, and the points in the Q‑Q plot largely follow the reference line $y=x$, together indicating that the model‑generated PDFs are well‑calibrated and reliably reflect the predictive uncertainty. Although the histogram exhibits a slight U‑shape (with slightly elevated tails) and the Q‑Q plot shows a subtle S-shaped trend—suggesting a mild under‑dispersion, i.e., a slight underestimation of the uncertainty that makes the predictions somewhat overconfident—this bias is small and does not compromise the reliability of the redshift confidence intervals provided by the model for the vast majority of galaxies. Overall, the good probabilistic calibration ensures that the model is suitable for subsequent cosmological statistical analyses.

To further investigate the impact of the bimodal distribution in our sample (dominated by the MGS at $z < 0.4$ and LRGs at $0.4 < z < 1.0$), we evaluated the model's performance on these two sub-populations separately. For the low-redshift (MGS-like) sample, ViT-MDNz achieved a $\sigma_{\rm NMAD}$ of 0.034 and $f_{\rm out}$ of 4.0\%; for the high-redshift (LRG-like) sample, it reached a $\sigma_{\rm NMAD}$ of 0.034 and $f_{\rm out}$ of 1.2\%.

The consistent accuracy across these intervals indicates that the model performs fine-grained redshift inference rather than simple population classification with prior assignment. If the model were only performing coarse classification, one would expect the correlation between $z_{\rm pred}$ and $z_{\rm spec}$ within each sub-interval to flatten or exhibit significantly higher dispersion. However, we observe a continuous and tight linear correlation in both regimes, confirming that the Vision Transformer effectively extracts morphological features (e.g., angular size and concentration) that evolve with redshift within different galaxy types.

Regarding probabilistic calibration, we also performed detailed diagnostics for the MGS-like ($z < 0.4$) and LRG-like ($0.4 < z < 1.0$) samples. Internal analysis shows that the Probability Integral Transform (PIT) distributions for both sub-samples remain nearly uniform, and their Q-Q plots exhibit the same slight S-shaped trend as the full sample. This cross-sample consistency demonstrates that the model maintains robust statistical behavior across different galaxy populations and redshift ranges. Consequently, the model exhibits well-calibrated performance for each galaxy sub-population.

\begin{figure} 
    \centering
    \includegraphics[width=0.5\textwidth]{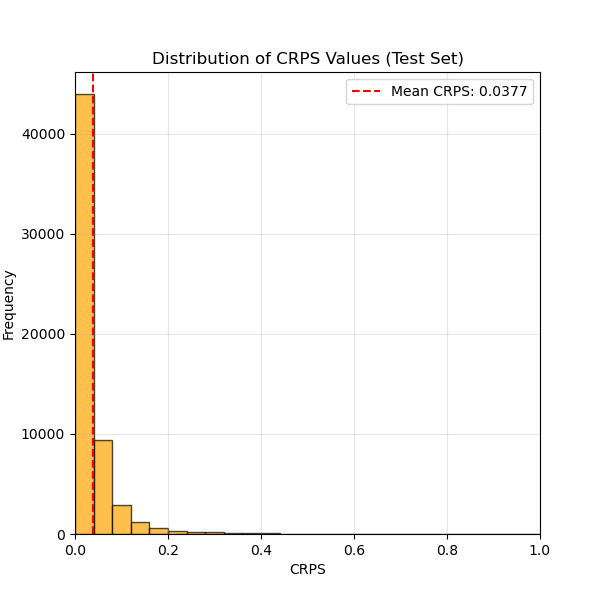} 
    \caption{ Distribution of CRPS values for the $r$-band test set. The red dashed line indicates the mean CRPS value.} 
    \label{fig:crps} 
\end{figure}

The accuracy of the probabilistic predictions was further evaluated using CRPS.
As shown in Fig.~\ref{fig:crps}, the distribution of CRPS values over the full $r$-band test set has a mean of about 0.0377. While there is no theoretical minimum for CRPS, this value is comparable to the typical scatter observed in point estimation ($\sigma_{\rm NMAD}$), suggesting that the predicted PDF is well-concentrated around the true redshift. This score serves as a baseline for single-band performance, which we later show improves as more color information is introduced.

\begin{figure} 
    \centering
    \includegraphics[width=\textwidth]{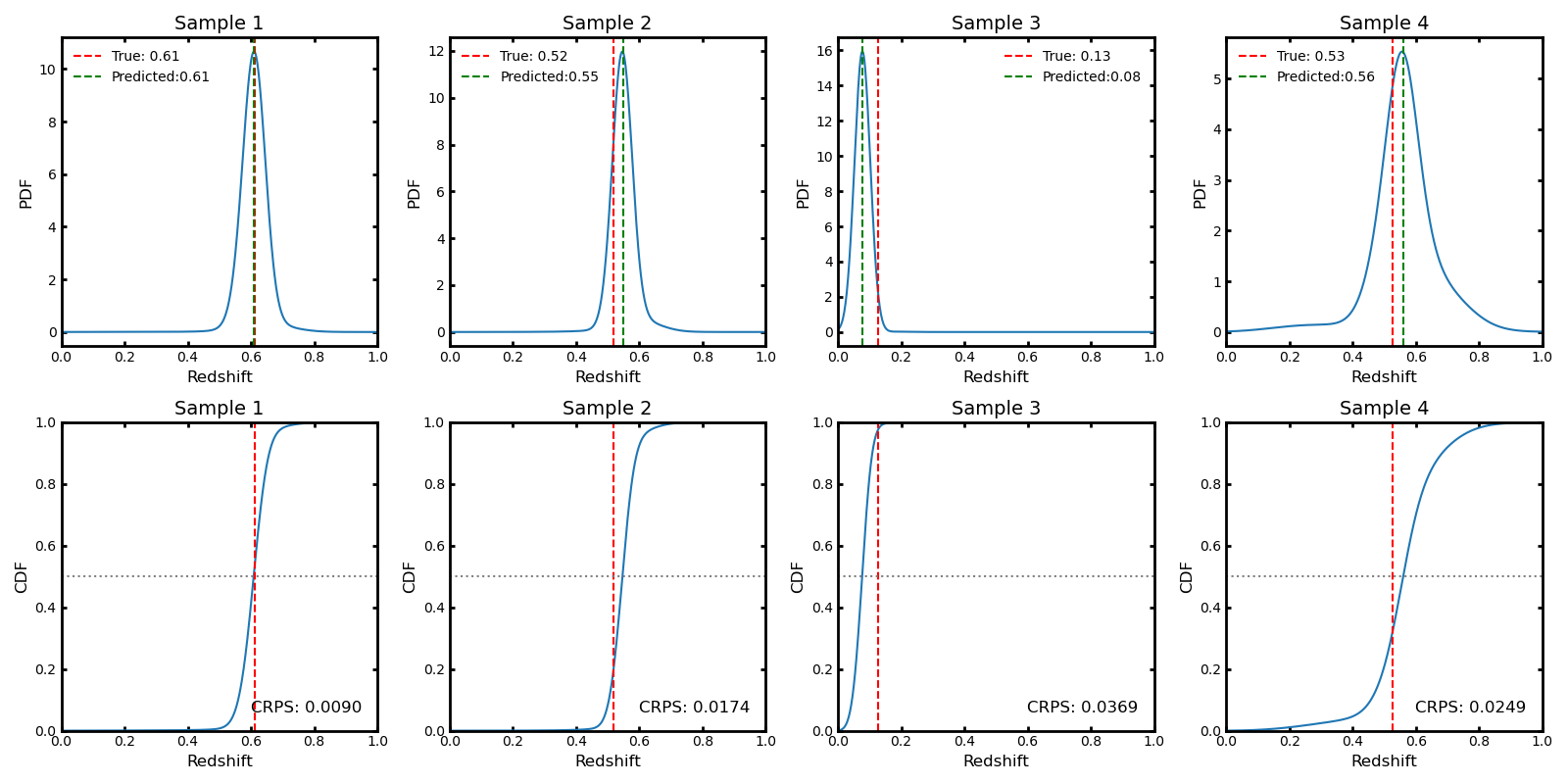} 
    \caption{Predicted redshift PDFs for four randomly selected galaxies from the $r$-band test set. The upper panel of each subplot shows the predicted PDF, and the lower panel shows the corresponding CDF. The red dashed line marks the true spectroscopic redshift, and the blue dashed line represents the point estimate derived from the predicted PDF.} 
    \label{fig:pdf_cdf} 
\end{figure}

Figure \ref{fig:pdf_cdf} displays the predicted PDFs and their corresponding CDFs for four randomly selected galaxies from the test set. In each subplot, the true spectroscopic redshift, marked by a red dashed line, lies within the high‑probability region of the predicted PDF. This demonstrates that the model not only provides accurate point estimates but also yields high‑quality, reliable uncertainty quantification. 

Finally, we address the potential redshift degeneracies inherent in single-band inference by examining the multimodality of the predicted PDFs. By identifying the local maxima in the $r$-band test set, we find that only approximately 0.7\% of the samples exhibit significant multimodal distributions (defined as having a secondary peak with at least 10\% of the primary peak's amplitude). This low fraction indicates that the high-resolution morphological features extracted by the ViT provide sufficiently unique priors to resolve most redshift ambiguities. For the vast majority of cases, the model expresses predictive uncertainty through the widening of a single PDF mode rather than through discrete, degenerate solutions.

\subsection{Multi-band Extension}

The ViT-MDNz model is also capable of handling multi-band images. To investigate the potential performance gain from additional color information, we progressively expanded the input from single-band to three bands. First, by concatenating images along the channel dimension, we constructed all possible dual-band combinations ($g+r$, $r+z$, $g+z$), resulting in 2‑channel inputs. We then fully concatenated the $g$, $r$, and $z$ bands to form a 3‑channel input that simulates a color image. Table \ref{tab:band_com} summarizes the performance results for these input configurations.

\begin{table}[ht]
    \centering
    \caption{Multi-band Input Performance Comparison (Baseline: single $r$-band)}
    \label{tab:band_com}
    \setlength{\tabcolsep}{8pt} 
    \renewcommand{\arraystretch}{1.2} 
    \begin{tabular}{ccccccc}
        \hline
        \textbf{
Input Configuration} & $\bm{\sigma_{\rm NMAD}}$ & $\bm{f_{\rm out} (\%)}$ & $\bm{bias}$ & \textbf{MSE} & \textbf{MAE} & \textbf{CPRS (mean)} \\
        \hline
        $r$-band & 0.034 & 2.6 & -0.0012 & 0.0075 & 0.051 & 0.0377 \\
        $g+z$ & 0.028 & 2.7 & -0.0046 & 0.0075 & 0.049 & 0.0356\\
         $g+r$ & 0.026 & 2.1 & -0.0021 & 0.0073 & 0.043 & 0.0332 \\      
        $r+z$ & 0.024 & 2.0 & -0.0022 & 0.0063 & 0.041 & 0.0300\\         
        $g+r+z$ &\textbf{0.020} &\textbf{1.5} &\textbf{-0.0026} &\textbf{0.0049} &\textbf{0.034} &\textbf{0.0244}\\
        \hline
    \end{tabular}
    \vspace{0.3cm}
\end{table}

The results in Table \ref{tab:band_com} demonstrate that incorporating multi-band information substantially enhances model performance. Almost all dual- and triple-band configurations outperform single-band inputs across both point-estimation and PDF evaluation metrics, with the three-band ($g+r+z$) combination achieving the best results. This indicates a strong complementarity between the color information provided by multiple bands and the morphological features captured from single-band images. ViT-MDNz effectively integrates these two types of information: it reduces $\sigma_{\rm NMAD}$ from 0.034 for the single $r$-band to 0.020 for the three-band input, lowers the outlier rate by approximately 43\% to 1.5\%, and significantly decreases the CRPS from 0.0377 for the $r$-band to 0.0244. These findings validate the effectiveness of the model architecture in fusing multi-band photometric data and spatial information extracted from the images, and demonstrate that ViT‑MDNz remains highly competitive when full multi-band data are available, making it a flexible approach adaptable to different data conditions.

\begin{figure} 
    \centering
    \includegraphics[width=0.8\textwidth]{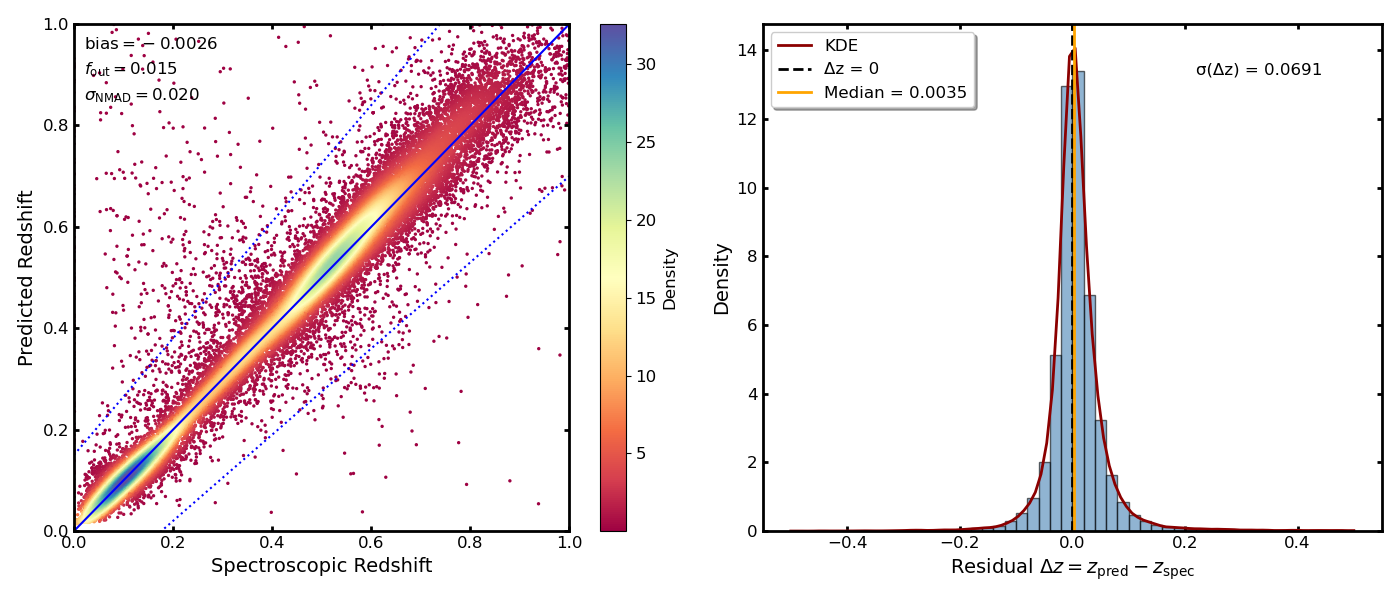} 
    \caption{(Left) Predicted redshift vs. spectroscopic redshift scatter plot for the three-band (g+r+z) input. (Right) Corresponding residual distribution. } 
    \label{fig:zsp_zph} 
\end{figure}

Figure \ref{fig:zsp_zph} visually presents the point redshift estimation results for the three-band input. Compared to the single-band ($r$‑band) results in Figure \ref{fig:sing_zsp_zph}, the predicted and true redshifts under the three-band input are more tightly clustered around the diagonal, and the scatter of the residuals is noticeably reduced. These observations indicate significant improvements in both the accuracy and consistency of the point redshift estimates.

\section{Summary and Discussion} \label{sec:summary}

This work addresses the heavy reliance of traditional photometric redshift methods on multi-band photometric data by proposing a new approach for redshift estimation under the extreme condition of only single-band images. We introduce ViT-MDNz, a novel deep learning architecture that, for the first time, integrates a vision transformer (ViT) with a mixture density network (MDN). This model is capable of learning redshift-related deep morphological features directly from single-band galaxy images and outputs a complete photometric redshift probability density function.


In experimental validation based on a large sample from the DESI Legacy Imaging Surveys, the ViT‑MDNz model demonstrates competitive performance under single-band settings within the redshift range of $z \lesssim 1$. Specifically, in the $r$-band, the model achieves a normalized median absolute deviation ($\sigma_{\rm NMAD}$) of 0.034 and an outlier fraction ($f_{\rm out}$) of 2.6\% for point estimates of redshift, indicating that relatively accurate single-band redshift estimation can be achieved using morphological information alone. Further evaluation based on probability integral transform (PIT) analysis and the continuous ranked probability score (CRPS) shows that the predicted redshift probability density functions align closely with the true distribution, confirming that the outputs are well calibrated and can reliably quantify predictive uncertainties, thereby meeting the requirements for subsequent cosmological statistical applications.

It should be noted that single-band images inherently contain limited information and cannot fully break the degeneracy between redshift, galaxy type, and luminosity, making precise redshift determination challenging. However, the feasibility demonstrated by this method does not stem from a breakthrough in physical principles but rather benefits from statistical priors learned from large-scale data. Under strictly constrained data conditions, this remains a worthwhile direction for in-depth exploration. Our findings are consistent with recent studies on next-generation survey data, such as the DeepDISC-Photoz framework for LSST \citep{2025OJAp....8E..40M}, which confirmed that deep learning models actively utilize pixel-level structural information beyond integrated photometry to refine redshift estimates.

In multi-band extension experiments, model performance improves significantly as more band information is incorporated. When using combined $g$, $r$, and $z$-band inputs, $\sigma_{\rm NMAD}$ further decreases to 0.021, and the outlier rate drops by nearly 43\% to 1.5\%, validating the effectiveness and flexibility of the ViT-MDN architecture in integrating color information with morphological features. This makes it a highly competitive solution even when complete multi-band data are available.

The contributions and significance of this work are reflected in three primary aspects. First, at the paradigm level, this study systematically demonstrates the feasibility of estimating redshifts directly from morphological features in single-band images. This challenges the conventional assumption that photometric redshifts must rely on multi-band color information, thereby opening a new methodological pathway for the field. Second, on the technical front, we explore the integration of the ViT and MDN as a specialized architecture for photometric redshift estimation. This combination leverages ViT global feature modeling to capture high-dimensional morphological cues that local operators might overlook, while utilizing MDN to provide a robust probabilistic framework to generate complete PDFs of redshift. By synergistically linking these two components, our approach facilitates a more nuanced characterization of the relationship between galaxy structure and cosmic distance. Third, in terms of practical application, our method offers a viable solution for galaxy samples with incomplete band coverage in upcoming large-scale surveys such as the Chinese Space Station Telescope (CSST; \citealt{2019ApJ...883..203G}), LSST, and Euclid. By reducing the discard rate of galaxies due to missing data, this approach enhances the completeness of redshift samples, mitigates selection biases from non-random exclusion, and assists in source prioritization for follow-up spectroscopic observations.

Although this study has yielded promising results, several limitations warrant further discussion. Firstly, the accuracy of single-band redshift estimation remains inferior to that of high-quality multi-band methods, a fundamental constraint stemming from the limited information content of morphological features, which cannot impose redshift constraints as strong as those provided by color information. Secondly, the decision-making mechanism through which the model learns redshift-related features from single-band images remains unclear, and its ``black-box'' nature currently hinders physical interpretability. Specifically, while the model achieves high predictive accuracy, it remains challenging to identify which specific morphological markers are being prioritized. Finally, the generalizability of the model for galaxies with high-redshift ( $z > 1$ ) and galaxies with peculiar morphological features is limited by the representativeness of the current training sample and requires further validation with larger and more diverse datasets.

To address these challenges, future research could focus on the following directions: leveraging observational images with higher signal-to-noise ratios and improved resolution to enhance data quality; developing more efficient lightweight ViT architectures; implementing explainable AI (XAI) techniques, such as attention visualization or saliency maps, to unveil the model's decision-making process; introducing physics-guided constraints that combine prior knowledge with data-driven approaches; and constructing a unified, multimodal, cross‑dataset redshift estimation framework to facilitate generalization and application across diverse datasets and tasks.

In summary, the ViT-MDN model proposed in this work not only provides a practical tool for addressing redshift estimation for "data-incomplete" galaxies but also challenges traditional assumptions at a methodological level, confirming the independent value of morphological information in redshift estimation. With the continuous accumulation of observational data and the ongoing evolution of deep learning architectures, morphology-based redshift estimation methods are expected to play a key role in probing deeper into the universe and constructing more complete cosmological samples.

\begin{acknowledgments}

Z.L. acknowledges the support from the National Natural Science Foundation of China (Grant No. 12573009) and the scientific research grants from the China Manned Space Project with Grand No. CMS-CSST-2025-A07 and CMS-CSST-2025-A05. S.Z. acknowledges support from the National Natural Science Foundation of China (Grant No. 12173026), the Program for Professor of Special Appointment (Eastern Scholar) at Shanghai Institutions of Higher Learning, and the Shuguang Program (23SG39) of the Shanghai Education Development Foundation and Shanghai Municipal Education Commission. This work is also supported by the National Natural Science Foundation of China under Grant No. 12141302. 

\end{acknowledgments}

\bibliography{ref}{}
\bibliographystyle{aasjournalv7}



\end{document}